\documentclass[10pt,journal,compsoc]{IEEEtran}

 \pdfoutput=1 
\ifCLASSOPTIONcompsoc
  \usepackage[nocompress]{cite}
\else
  \usepackage{cite}
\fi

\usepackage[pdftex]{graphicx}
\usepackage{verbatim}
\usepackage{subfigure}
\usepackage{booktabs}
\usepackage{tabularx}
\usepackage{multirow}
\usepackage{booktabs}
\usepackage[T1]{fontenc}
\usepackage{ragged2e}

\usepackage[justification=centering]{caption}

\begin{document}

\title{CodeGen-Test:An Automatic Code Generation Model Integrating Program Test Information}

\author{ZHONG Maosheng,
        LIU Gen,
        LI Hongwei*,
        KUANG Jiangling,
        ZENG Jinshan,
        WANG Mingwen*
        
\IEEEcompsocitemizethanks{\IEEEcompsocthanksitem ZHONG Maosheng,
        LIU Gen,
        LI Hongwei,
        KUANG Jiangling,
        ZENG Jinshan,
        and WANG MingWen are with the School of Computer and Information Engineering,  Jiangxi Normal University, Nanchang,China,
330022.\protect\\
E-mail: zhongmaosheng@sina.com,\\ \{sun,lihongwei,jianglingk,jinshanzeng,  mwwang\}@jxnu.edu.cn 
}
\thanks{}
}

\markboth{}%
{Shell \MakeLowercase{\textit{et al.}}: Bare Demo of IEEEtran.cls for Computer Society Journals}

\IEEEtitleabstractindextext{
\begin{abstract}
\justifying
Automatic code generation is to generate the program code according to the given natural language description. The current mainstream approach uses neural networks to encode natural language descriptions, and output abstract syntax trees (AST) at the decoder, then convert the AST into program code. While the generated code largely conforms to specific syntax rules, two problems are still ignored. One is missing program testing, an essential step in the process of complete code implementation; the other is only focusing on the syntax compliance of the generated code, while ignoring the more important program functional requirements.
The paper proposes a CodeGen-Test model, which adds program testing steps and incorporates program testing information to iteratively generate code that meets the functional requirements of the program, thereby improving the quality of code generation.
At the same time, the paper  proposes a new evaluation metric, test accuracy (Test-Acc), which represents the proportion of passing program test in generated code. Different from the previous evaluation metric, which only evaluates the quality of code generation from the perspective of character similarity, the Test-Acc can evaluate the quality of code generation from the Program functions. Moreover, the paper evaluates the CodeGen-test model on a python data set "hearthstone legend". The experimental results show the proposed method can effectively improve the quality of generated code. Compared with the existing optimal model, CodeGen-Test model improves the Bleu value by 0.2\%, Rouge-L value by 0.3\% and Test-Acc by 6\%.
\end{abstract}

}

\maketitle

\IEEEdisplaynontitleabstractindextext

\IEEEpeerreviewmaketitle

\IEEEraisesectionheading{\section{Introduction}\label{sec:introduction}}

%
%
%
%

 Automatic code generation is to generate the program code according to the given natural language description. The early approaches are mainly based on templates\cite{zettlemoyer2007online,zettlemoyer2012learning,kushman2013using,wang2014morpho}. In recent years, with the development of neural machine translation    (NMT), sequence-to-sequence (Seq2Seq) model\cite{sutskever2014sequence,cho2014learning,gehring2017convolutional,luong2015multi} have been used for automatic code generation\cite{hayati_retrieval-based_2018,sun_grammar-based_2019,sun_treegen_2019,lyu_embedding_2021}. 
 However, in the early stage, only simple code vocabulary sequence can be generated\cite{mou_end--end_2015,ling_latent_2016}, which can not meet the requirements of correct code grammar. So researchers are exploring how to generate code that follows the underlying syntax. For example, Dong et al.\cite{dong_language_2016} proposed the Seq2Tree model by generating multi-layer tree to meet the structural requirements of the code; Yin et al.\cite{yin_syntactic_2017}  proposed a method of generating abstract syntax tree based on sequence to sequence model to make the code meet the syntax requirements; Rabinovich et al.\cite{rabinovich_abstract_2017} chose
different decoders according to different nodes of the generated abstract syntax tree to improve the performance of the model; Li et al.\cite{wei2019code} propose code generation as a dual task of code summarization;Sun et al.\cite{sun_grammar-based_2019,sun_treegen_2019}  used convolutional neural network 
\cite{krizhevsky2012imagenet} and Transformer \cite{vaswani_attention_2017} to improve the performance of generating abstract syntax tree.  With the development of pre-training model\cite{devlin2018bert,lewis2019bart, radford2019language} , it has also been proposed to help automatic code generation by pre-training models\cite{feng2020codebert, guo2020graphcodebert,paik2021improving,xu2020incorporating}.

At present, the method based on abstract syntax tree is used to improve the grammatical correctness of the generated code, and the evaluation metric such as BLEU \cite{papineni_bleu_2002}  and Accuracy  have been significantly improved. However, high-quality code must not only satisfy the correctness of the code syntax, but also satisfy the correctness of the code function. Therefore, the code testing phase in the code generation process is indispensable, which can further ensure the quality of code generation. In order to solve this problem, this paper adds a program test step in the code generation process, by quantifying the performance of the generated code on the functional correctness, and combining the program test information to further guide the model to generate code that can meet both the grammatical requirements and the functional requirements. 

The contributions of this paper are as follows:

1. We propose to add the program test step and  incorporate test information  in the process of automatic code generation , then generate code by an iterative manner. This provides a novel idea and direction for future research on automatic code generation.

2. We propose a new evaluation metric, Test-Acc, compared with the existing evaluation metric, it can evaluate the generated code in terms of code function.

Section 2 and 3 of this paper introduces the related work and  knowledge of automatic code generation respectively ; Section 4 introduces the CodeGen-Test model ; Section 5 introduces the experimental design and experimental results, and analyzes the model; Section 6 summarizes the work of this paper and discusses the prospect of future work.

\section{Related work}

\subsection{Automatic code generation model based on Neural Network}

In view of the excellent performance of neural network sequence to sequence  framework \cite{sutskever_sequence_2014} in the field of machine translation, researchers apply this method to the task of automatic code generation. Mou et al. \cite{mou_end--end_2015} proposed an end-to-end program generation model based on recurrent neural network, which uses natural language as input to generate the corresponding code character by character. In order to alleviate the generation problem of low-frequency words in the code generation task, Ling et al.\cite{ling_latent_2016} proposed a code generation model integrating pointer network, by considering the direct use of words in the input . The model copies characters or words from the input source directly as a part of the output code. In addition, considering the important role of code API information, Chen et al.\cite{lyu_embedding_2021} proposed to fuse code API information, model the dependency relationship between API methods as API dependency graph (ADG), and embed the graph into the neural network sequence model.

The neural network-based code generation model generally adopts a sequence-to-sequence framework. By combining different characteristics of the code, such as code variables, code API information, etc, the performance is significantly improved. At the same time, the researchers noticed that the grammatical information of the code is ignored during the automatic code generation process, so most of the generated code cannot run normally. To this end, Dong et al.\cite{dong_language_2016} proposed to use a tree structure to represent the code level, and used a recurrent neural network to generate the code hierarchy by outputting sequences and trees. 
Jiang et al. \cite{jiang2021exploring} propose to add a context-based branch selector to the Seq2Tree model and optimize the selector by reinforcement learning to dynamically determine the optimal branch expansion order for multibranch nodes. Xie et al. \cite{xie2021improving} propose a model to improving Tree-Structured Decoder training via mutual learning. Although these method takes into account the code structure level, the generated code still has grammatical problems. 

\subsection{Automatic code generation model integrating abstract syntax tree}

Abstract syntax tree is an abstract representation of the syntax structure of source code, which can be transformed into source code. It represents the syntax structure of code in the form of tree. Each node on the tree represents a structure in source code.

Researchers conduct in-depth exploration on how to fully integrate code abstract syntax trees in neural networks. Yin et al.\cite{yin_syntactic_2017} proposed to use the abstract syntax tree as the intermediate state of the code generation process. The model divides the code generation task into two steps: first generate the abstract syntax tree from the natural language intent, second  convert the abstract syntax tree to code. The method  saves the model from learning the underlying syntax of the code from limited training data, allowing the model to focus on the generation of abstract syntax tree nodes. In order to improve the correct rate of generating the sub-nodes of the abstract syntax tree, Rabinovich et al.\cite{rabinovich_abstract_2017} proposed that the decoder of the model can dynamically select the corresponding decoder structure according to the node category of the abstract syntax tree. Hayati et al.\cite{hayati_retrieval-based_2018} believe that existing code examples can be used in the code generation process, and a search-based code generation model is proposed on the basis of fusion of abstract syntax trees. The text similarity method is used to retrieve sentence-code pairs that are similar to the input sentence, and subtrees are extracted from the abstract syntax tree of the code, and the probability of generating the subtree is increased in the process of generating the abstract syntax tree.

Compared with sequence-to-sequence models such as RNN\cite{williams1989learning} and LSTM\cite{hochreiter_long_1997}, convolutional neural networks (CNN) can effectively capture the features of different regions through sliding windows. Based on the fusion of abstract syntax trees, Sun et al.\cite{sun_grammar-based_2019} proposed a code generation model based on convolutional neural networks, using convolutional neural networks to capture the structural information of the tree. Later, the author proposed a TreeGen\cite{sun_treegen_2019} model that employs Transfromer\cite{vaswani_attention_2017} architecture to capture the characteristics of sentence-length dependency, and further improved the performance.

The integration of abstract syntax trees effectively improves the problem of incorrect syntax in generated code and further advances the field of automatic code generation.

\section{Basic knowledge}

The method proposed in this paper is based on the generation of code abstract syntax tree.So this section introduces the basic knowledge of abstract syntax trees(AST) and evaluation metric commonly used in code generation tasks.

\subsection{Abstract syntax tree of code}

\begin{figure}[htbp]
    \centering
    \includegraphics[width=3in]{ 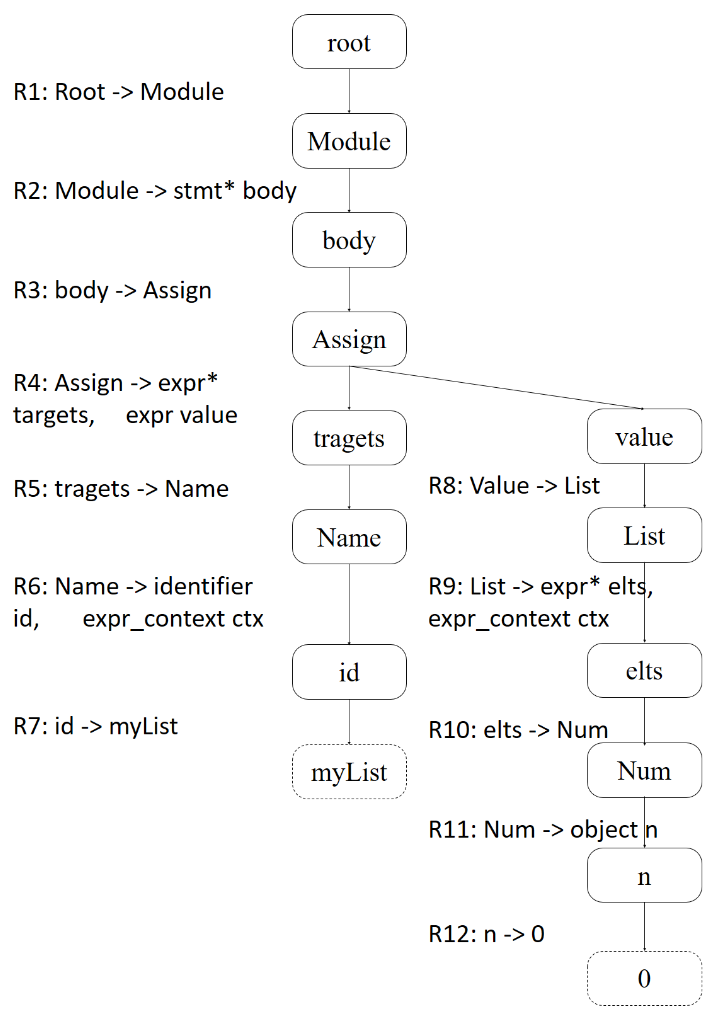}
    \caption{Abstract syntax tree corresponding to Python code "mylist = [0]"}
    \label{fig:AST.png}
\end{figure}

The abstract syntax tree is an intermediate state in the code generation process. We can expand the tree nodes by predicting the abstract syntax tree rules step by step until the complete abstract syntax tree is generated, and finally convert it to code. For example, to generate the code "mylist = [0]", it can convert it to an abstract syntax tree as shown in figure \ref{fig:AST.png}. In figure \ref{fig:AST.png}, the solid wireframe represents non-terminal nodes, the dotted line box represents terminal nodes, and R1 to R10 on the left side of the figure represent rules. For example, "assign - > expr *targets, expr values" represents that the current node generates "targets" and "value" nodes through this rule, which corresponds to the assignment statement of the code. The abstract syntax tree generated in the graph can be then directly transformed into  code with correct syntax.This approach effectively improves the problem of incorrect syntax of the generated code.

\subsection{Related evaluation metric for code generation task \label{sec:metric}}

Most of the existing models of automatic code generation task use Bleu  value\cite{papineni_bleu_2002}, Accuracy  and Acc+ \cite{sun_grammar-based_2019} to evaluate the performance of the model.

BLEU value is an metric for automatically evaluating translation performance in the field of machine translation. In the field of automatic code generation, the performance of generated codes is measured by calculating the closeness of the generated code and the target code in n-grams. But BLEU is not entirely applicable to the field of automatic code generation, because it ignores some characteristics of the code generation tasks. For example,  code with the following characteristics, the BLEU value cannot be a reasonable and effective evaluation model:

1. Change of variable name  does not affect code function. For example, all variables a1 in the function are replaced by the variables a2, where a2 is the name of the variable that has never been used. Although the code is completely correct, the Bleu value is decreased because of the variable changed.

2. Change of statement sequence   does not affect code function. For example, the code "a = 1; b = 1; c = a + b;" is changed to "b = 1;a = 1; c = a + b;".Although the code is completely correct, the Bleu value is also decreased because the code sequence is changed.

3. Different logic implementation methods that do not affect the code function, i.e., different codes realize the same function.

Accuracy (Acc) is the most stringent metric. It represents the ratio of samples in the generated code that are exactly the same as the target code to the total samples. However, it is the same as the BLEU value, which tends to give high scores for completely consistent codes. For codes with the above three characteristics, even if the codes are completely correct, the accuracy will be given low scores.

Manual correction accuracy rate (Acc+) is based on the accuracy rate, which incorporates the above-mentioned feature 1 "change of variable name   does not affect the code function" into the calculation accuracy rate. Although it compensates for the shortcomings of accuracy to a certain extent, it still cannot reasonably evaluate the performance of the model under the other two characteristics.

\section{CodeGen-Test model}
Inspired by the complete code development process, we propose the  CodeGen-Test model, which adds program test step in the process of code generation, measures the performance of generated code on  function with Test-Acc Metric , and further guides the model to generate higher quality code combined with program test information.

\begin{figure}[htbp]
    \centering
    \includegraphics[width=2in]{ 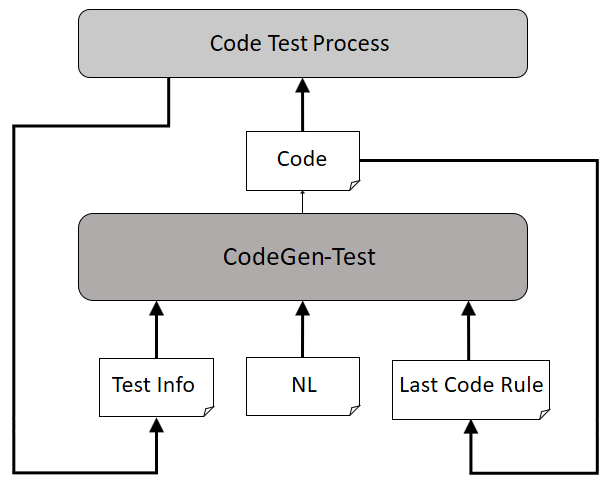}
    \caption{Model flowchart}
    \label{fig:2-model-mini.png}
\end{figure}

The model is an iterative mode, and N represents the number of iterations. As shown in Figure \ref{fig:2-model-mini.png}, in the first round of iteration, the model  generates  code from the input of natural language description (NL), and then converts the code  to  abstract syntax tree to obtain rule sequence (Last Code Rule) of the abstract syntax tree. Meanwhile, the generated code is input into the code test process to obtain the Test Information (Test-Info).In the second iteration, input the Test-Info, Last Code Rule and NL obtained in the previous round into the model, generate the code again, and then obtain the Last Code Rule and Test-Info according to the newly generated code. The model performs 
multiple rounds iteration . In the first round of iteration, only NL is input, and both Test-Info and Last Code Rule are empty.

\begin{figure*}[htbp]
    \centering
    \includegraphics[width=5in]{ 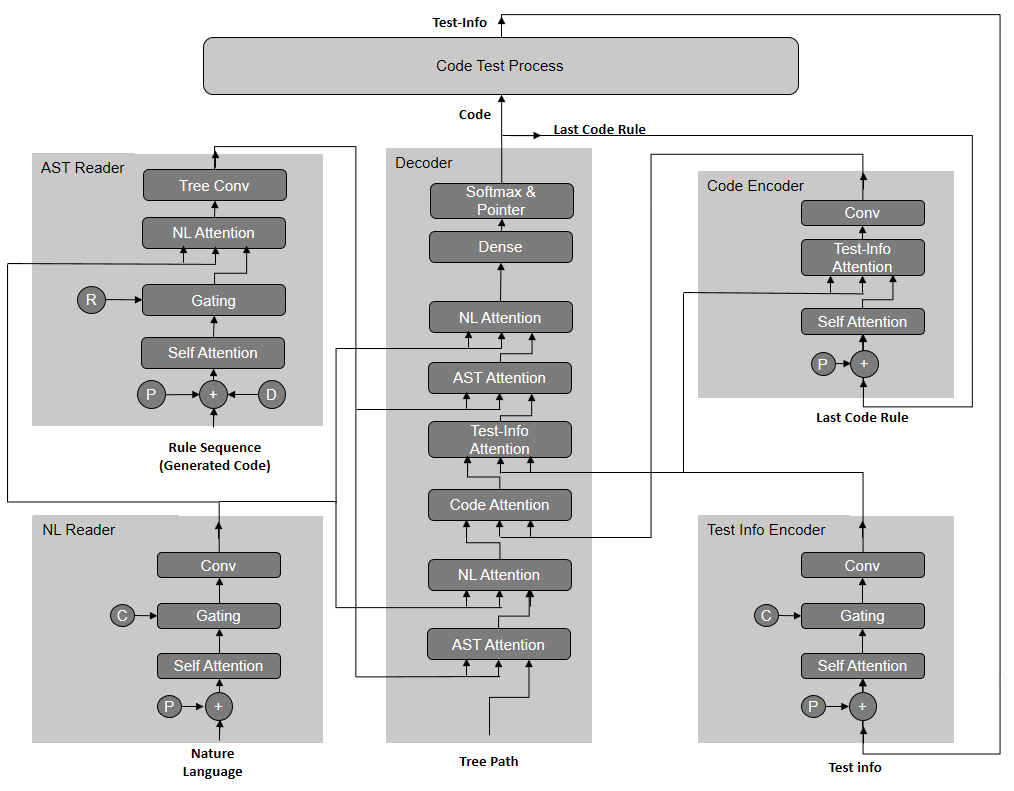}
    \caption{CodeGen-Test model }
    \label{fig:3-model.png}
\end{figure*}

The model flowchart is shown in Figure \ref{fig:3-model.png}. Based on TreeGen \cite{sun_treegen_2019}, this model adds Test-Info Encoder and Code Encoder. The information output by each encoder is integrated in the decoder to generate code by generating abstract syntax tree. In the generation process, pointer network \cite{see_get_2017} is used to fill the terminal node of abstract syntax tree.

The CodeGen-Test model includes five modules: natural language encoder (NL Reader), abstract syntax tree encoder (AST Reader), Test-Info Encoder, Code Encoder and Decoder. Same as in TreeGen\cite{sun_treegen_2019}, NL Reader encode natural language descriptions (NL), and AST Reader encode the structure of the partial AST that has generated , Decoder is responsible for predicting the output. The following describes the Test-Info Encoder and Code Encoder modules in this paper.

\subsection{Test-Info Encoder}\label{section:Test-Info Encoder}

 The fusion of test information helps the model learn the shortcomings of the existing code and improve it in the next generation process. The Test-Info Encoder is responsible for receiving Test Information (Test-Info). The encoder is composed of blocks, each block contains three different sub layers (self-attention, gating mechanism and convolution ). Between each two sub layers, residual connection \cite{he2016deep} is adopted , and layer normalization \cite{ba_layer_2016} is carried out.

For an test information , we first tokenize it into tokens  ${n_1},{n_2}, \cdots ,{{\rm{n}}_{\rm{L}}}$ , where  $L$ is the length. Each token  ${n_i}$ is split into characters $c_1^{({n_i})},c_2^{({n_i})}, \cdots ,c_S^{({n_i})}$, where $S$ is the number of characters in the ${n_i}$. All the tokens and characters are represented as vector ${{\bf{n}}_{\bf{1}}}{\bf{,}}{{\bf{n}}_{\bf{2}}}{\bf{,}} \cdots {\bf{,}}{{\bf{n}}_{\bf{L}}}$ and ${\bf{c}}_1^{({n_i})},{\bf{c}}_2^{({n_i})}, \cdots ,{\bf{c}}_S^{({n_i})}$ by word embedding. For example, for test information "assertionerror: 3 != 1",${n_1},{n_2}, \cdots ,{{\rm{n}}_6}$ will be tokenize  into  'assertionerror', ':', '3', '!', '=', '1'. Each token will be split into characters for the gating mechanism below. For example, ${n_1}$ token "assertionerror" is split into characters $c_1^{({n_1})},c_2^{({n_1})}, \cdots ,c_{14}^{({n_1})}$, 'a','s','s', 'e', 'r','t ',' i ',' o ',' n ',' e ',' r ',' r ',' o 'and' r '.

\textbf{Self Attention} Attention mechanism can effectively alleviate the problem of long-distance dependence in sequence to sequence tasks. The self attention layer of the model follows the Transformer \cite{vaswani_attention_2017} architecture and uses the method  of Dehghani\cite{dehghani_universal_2018}. to embed the ${\rm{i}}$th position of the word in the ${\rm{b}}$th Transformer block as follows:
\begin{equation}
{{\rm{p}}_{b,i}}[2j] = \sin ((i + b)/({10000^{2j/d}}))
\label{equ:1}
\end{equation}

\begin{equation}
{{\rm{p}}_{b,i}}[2j + 1] = \cos ((i + b)/({10000^{2j/d}}))
\label{equ:2}
\end{equation}

where, $d$ represents the dimension of word embedding. Then, the latent semantic information is learned through multi head attention to generate a matrix:

\begin{equation}
Y^{(self)} = concat(head_1, \cdots ,head_h){W_h}
\label{equ:3}
\end{equation}
where, $h$ represents the number of multiple heads and $W_h$ is the weight matrix. The self attention mechanism is applied to each $head_t$, and the calculation equation is as follows:

\begin{equation}
head_t = softmax(\frac{{Q{K^ \top }}}{{\sqrt {{d_k}} }})V
\label{equ:4}
\end{equation}
where, ${d_k} = d/H$, $Q,K,V$ calculated by the following equation,

\begin{equation}
[Q,K,V] = {[{x_1}, \cdots ,{x_L}]^ \top }[{W_Q},{W_K},{W_V}]
\label{equ:5}
\end{equation}

where ${W_Q},{W_K},{W_V} \in {^{L \times d}}$ are the model parameters. ${x_i}$ is the input of this coding block, it is the vector sum of word embedding and position embedding in first block, that is ${{\bf{n}}_{\bf{i}}} + {p_{1,i}}$, and it is the vector sum of the output of the previous coding block and the position embedding of the block in other blocks.

\textbf{Gating Mechanism} After calculating self attention mechanism, the model adopts gating mechanism to further combine the information embedded in characters. As in the above example, the word "AssertionError" is split into characters, that is, 'A','s','s', 'e', 'r','t ',' i ',' o ',' n ',' e ',' r ',' r ',' r ',' o 'and' r ', which are embedded as characters ${\bf{c}}_1^{({n_1})},{\bf{c}}_2^{({n_1})}, \cdots ,{\bf{c}}_{14}^{({n_1})}$.

The vocabulary is represented by character embedding through the full connection layer.

\begin{equation}
{\bf{n}}_i^{(c)}{\rm{ = }}{{\rm{W}}^{{\rm{(c)}}}}[{\bf{c}}_1^{({n_i})},{\bf{c}}_2^{({n_i})}, \cdots ,{\bf{c}}_M^{({n_i})}]
\label{equ:6}
\end{equation}

For the  $i$th word, the output of the upper layer ${\rm{y}}_i^{{\rm{(self)}}}$ is linearly transformed to obtain the control vector ${q_i}$ and $k_i^{(y)}$,
\begin{equation}
q_i = {W_{1}}{\rm{y}}_i^{{\rm{(self)}}}
\end{equation}

\begin{equation}
k_i^{(y)} = {W_{2}}{\rm{y}}_i^{{\rm{(self)}}}
\end{equation}

Obtained $k_i^{(c)}$ from ${\bf{n}}_i^{(c)}$ in equation \ref{equ:6} by linear transformation,
\begin{equation}
k_i^{(c)} = {W_{3}}{\bf{n}}_i^{(c)}
\end{equation}
$\alpha _{i,t}^{(y)},\alpha _{i,t}^{(c)}$ are calculated as follows:
\begin{equation}
[\alpha _{i,t}^{(y)},\alpha _{i,t}^{(c)}] = softmax{\rm{\{ q}}_i^ \top k_i^{(y)}{\rm{,q}}_i^ \top k_i^{(c)}{\rm{\} }}
\end{equation}
Similarly, the upper layer feature $v_i^{(y)}$ and character embedding feature $v_i^{(c)}$ are obtained from ${\rm{y}}_i^{{\rm{(self)}}}$ and ${\bf{n}}_i^{(c)}$ by   linear transformation respectively,
\begin{equation}
v_i^{(y)} = {W_{4}}{\rm{y}}_i^{{\rm{(self)}}}
\end{equation}
\begin{equation}
v_i^{(c)} = {W_{5}}{\bf{n}}_i^{(c)}
\end{equation}

$\alpha _{i,t}^{(y)},\alpha _{i,t}^{(c)}$ for weighted $v_i^{(y)}$ and $v_i^{(c)}$,respectively.
\begin{equation}
{h_{i,t}} = [\alpha _{i,t}^{(y)}v_i^{(y)} + \alpha _{i,t}^{(c)}v_i^{(c)}]
\end{equation}
Similar to an arithmetic process as in equation 3, the output of the gating mechanism is ${Y^{(gate)}}$.

\textbf{Convolution} The convolution neural network can effectively capture the features of different regions through the sliding window. The model applies the convolution layer to the output of the gating mechanism layer $y_1^{(gate)}, \cdots ,y_L^{(gate)}$, to extract the local features $y_1^{({\rm{conv}},l)}, \cdots ,y_L^{({\rm{conv}},l)}$ around each word, where  $l$ represents the number of layers of the convolution layer. $y_i^{({\rm{conv}},l)}$ is calculated by the following equation:
\begin{equation}
y_i^{(conv,l)} = {W^{(conv,l)}}[y_{i - w}^{(conv,l - 1)}; \cdots ;y_{i + w}^{(conv,l - 1)}]
\label{equ:14}
\end{equation}
where, ${W^{(conv,l)}}$ is the weight of convolution kernel,$w = (k - 1)/2$, $k$ represents the size of convolution window. In particular, $y_i^{(conv,0)}$ represents the output of the gating mechanism layer $y_i^{(gate)}$.

To sum up, the Test-Info Encoder encodes the test information into $y_1^{(Test)},y_2^{(Test)}, \cdots y_L^{(Test)}$ after being processed by the self attention mechanism layer, gating mechanism layer and convolution layer.

\subsection{Code Encoder}
The code has a strong internal relationship with its test information. In the section \ref{section:Test-Info Encoder}, the feature of the test information has been extracted. In order to enable the model to learn the internal relationship between the code generated in the last round and its corresponding test information, the model constructs a code encoder.

We convert the code of last round generated to an abstract syntax tree to obtain the rule sequence . Then take rule sequence as input of code encoder. Suppose that the code generated in the last round is the code "mylist = [0]" shown in Figure \ref{fig:AST.png},  we convert the code to the abstract syntax tree shown in the figure  \ref{fig:AST.png}, then  obtain the rule sequences R1, R2,..., R12 by Pre-order traversal the abstract syntax tree, and take the R1 to R12 rule sequences as  the input of code encoder. 

The rule sequence of the abstract syntax tree generated in the previous round is express as ${r_1},{r_2}, \cdots ,{r_p}$, where $p$ represents the length of the sequence. Then these rules are embedded and expressed as vectors ${{\bf{r}}_1},{{\bf{r}}_2}, \cdots ,{{\bf{r}}_p}$.

\textbf{Self Attention} We also embed the position as in equation \ref{equ:1},\ref{equ:2}, then use the self attention mechanism with different weights as equation \ref{equ:3},\ref{equ:4},\ref{equ:5}, and finally output the feature vector $y_1^{({\rm{Code - self}})},y_2^{({\rm{Code - self}})}, \cdots y_L^{({\rm{Code - self}})}$.

\textbf{Test-Info Attention} In order to make the model better learn the relationship between the test information and the last round of generated code, the model integrates the vector representation extracted by the Test-Info Encoder in the test information attention layer.
The test information attention layer uses a multi head attention mechanism similar to the decoder of transformer \cite{vaswani_attention_2017}, sets up a multi head attention mechanism for fusing test information, and outputs feature $y_1^{({\rm{Code - att}})},y_2^{({\rm{Code - att}})}, \cdots y_L^{({\rm{Code - att}})}$ .

\textbf{Convolution} Finally, the model, as shown in equation \ref{equ:14}, uses convolution layers with different weights to learn the characteristics of different regions, and finally outputs $y_1^{({\rm{Code}})},y_2^{({\rm{Code}})}, \cdots y_L^{({\rm{Code}})}$.

\subsection{Decoder}
Decoder is responsible for fusing the information of each part and predicting the current rules of abstract syntax tree. On the basis of TreeGen \cite{sun_treegen_2019} model decoder, this decoder adds abstract syntax tree attention layer (AST attention) and natural language description attention layer (NL attention), which are respectively responsible for fusing the output of test information encoder and code encoder.

We take the tree path as the input of  Decoder. Tree Path is a path from the root node to the currently node that to be expanded in the abstract syntax tree of Generated code. For example, in Figure \ref{fig:AST.png}, assuming that the "assign" node needs to be expanded at present, the Tree Path is the "root", "module", "body" and "assign" nodes. The output of these nodes through a full connection layer is written as $q_i^{(path)}$.

We then apply  AST Attention and NL Attention layers to integrate the outputs
of the AST reader and the NL reader.  In each layer, the  $K$ and $V$ are  the output of the  corresponding reader and  ${\rm{Q}}$ is the  output of  the previous layer.

Similarly, the output of Test-Info Encoder and Code Encoder are integrated into the Test-Info  Attention layer and Code attention layer respectively . 

And apply  AST Attention and NL Attention with different parameters to output again. 
Then predicted the rule of  current nodes that need to expand by full connection layer. In the prediction process, the finger network \cite{see_get_2017} is also used to fill the terminal nodes of the abstract syntax tree.

\section{Evaluation}
\subsection{Dataset}
\begin{figure}[htbp]
    \centering
    \includegraphics[width=3.5in]{ 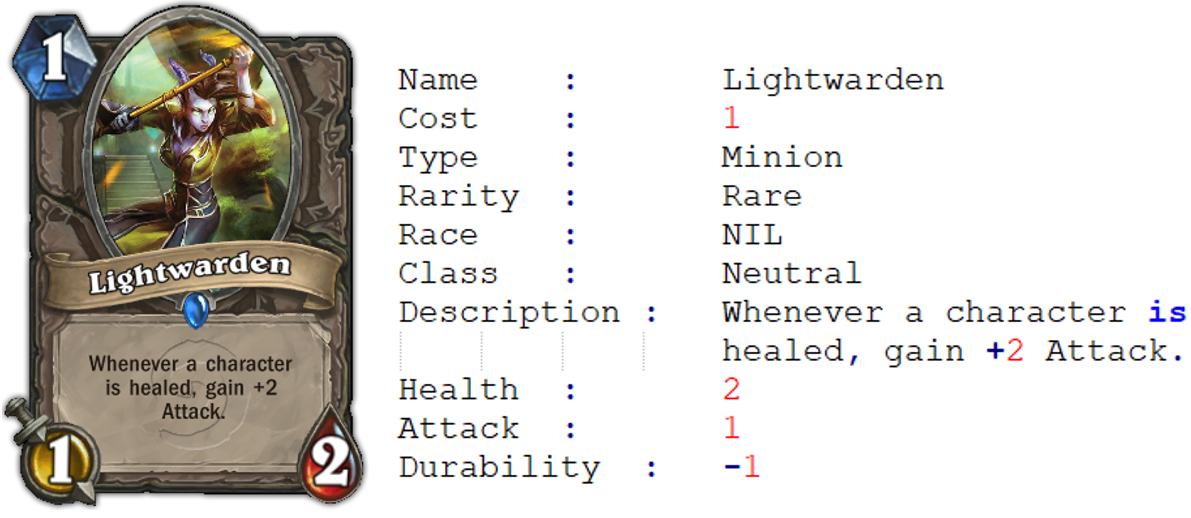}
    \caption{Cards and natural language descriptions}
    \label{fig:card}
\end{figure}

We  evaluated our approach on the "HearthStone" benchmark\cite{ling_latent_2016}, which  is a public benchmark dataset collected from  card game. It was produced by Blizzard Entertainment, a famous game production and distribution company. The data set contains Python code to implement 665 different cards, in which the sizes of training set, verification set and test set are 533, 66 and 66 respectively. Each card consists of a semi-structured description and a python program. This description is attached with some attributes, such as card name, card type, and natural language description of card function. As shown in Figure \ref{fig:card}, the natural language description part of the card and the correct code in Figure \ref{fig:a-correct-code} constitute a sample of the data set.

\subsection{Code Test Process and Test Information}
\begin{figure}[htbp]
    \centering
    \includegraphics[width=2in]{ 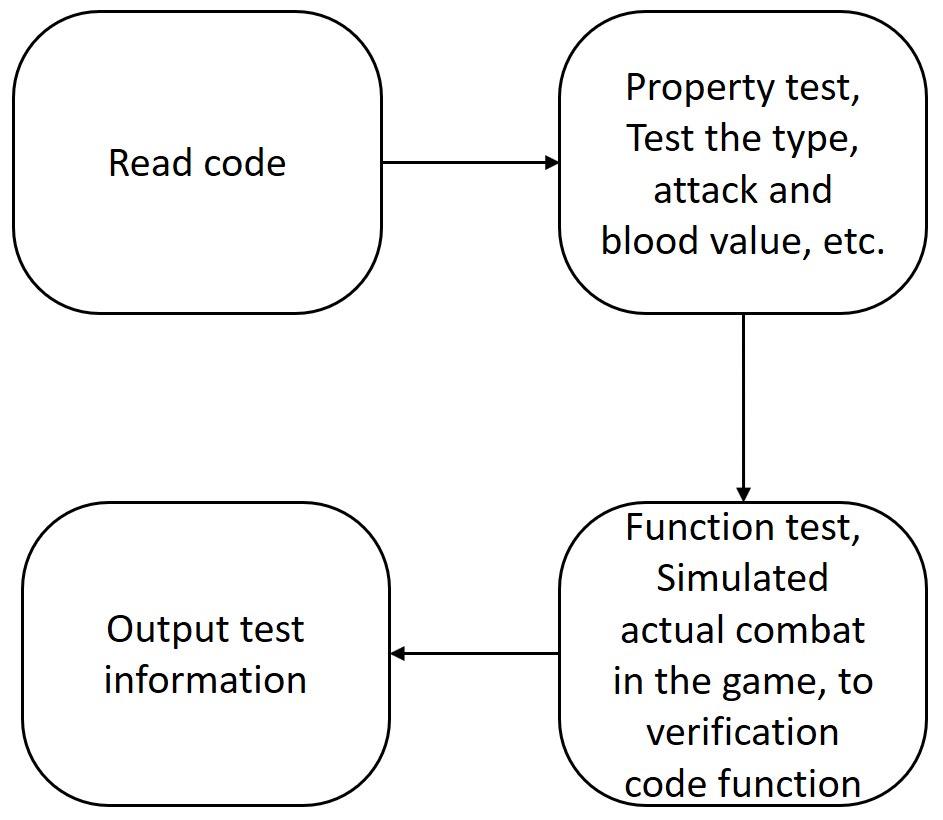}
    \caption{Code Test process Flowchart}
    \label{fig:test-flow}
\end{figure}
\begin{figure}
\centering
\subfigure[Error Code]{
    \begin{minipage}[b]{0.5\textwidth}
    \includegraphics[width=1\textwidth]{ 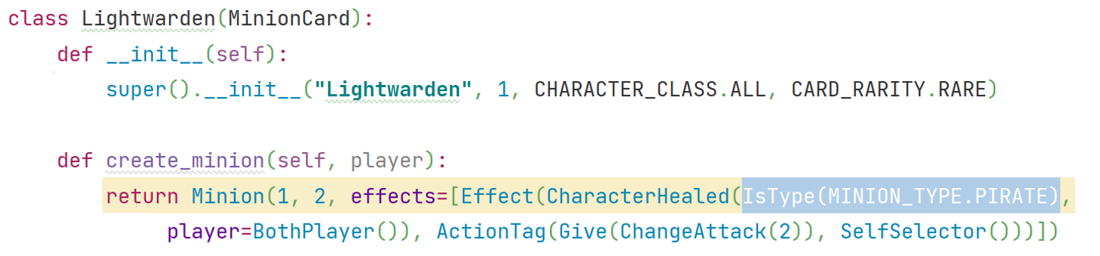}
    \label{fig:a-flase-code}
    \end{minipage}
}

\subfigure[Correct Code]{
    \begin{minipage}[b]{0.5\textwidth}
    \includegraphics[width=1\textwidth]{ 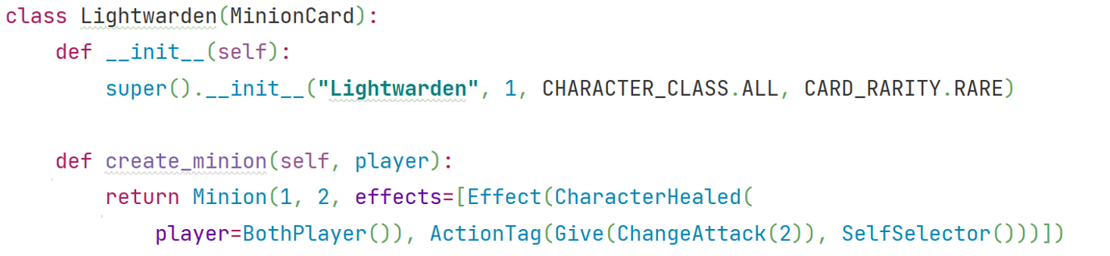}
    \label{fig:a-correct-code}
    \end{minipage}
}

\caption{Examples of error codes and correct codes} \label{fig:1}

\end{figure}
\begin{figure}[htbp]
    \centering
    \includegraphics[width=3.5in]{ 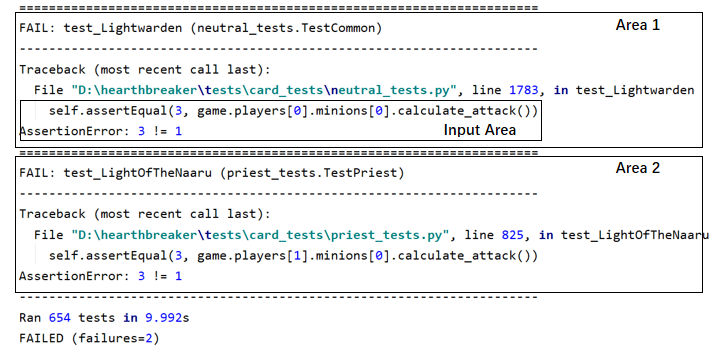}
    \caption{Test Information Example}
    \label{fig:test-info}
\end{figure}

\textbf{Code Test Process}
We uses the card test unit in the "hearthstone" open source simulator \footnote{https://github.com/danielyule/hearthbreaker} to test the generated code. The brief flow of card test unit is shown in Figure \ref{fig:test-flow}. After reading the code, the test unit first compares the attribute values in the code with the corresponding values in the system. If the comparison results are inconsistent, the information will be recorded. Then, the test unit tests the function of the code, inputs the code into the game simulator, simulates several rounds of actual combat, and checks whether the expected function is achieved. The test unit will return the test information shown in Figure \ref{fig:test-info} for the error code shown in Figure \ref{fig:a-flase-code}.

\textbf{Test Information}
Figure \ref{fig:test-info} is the result of testing the code shown in Figure \ref{fig:a-flase-code}, and it contains two errors (Area 1 and 2 in Figure \ref{fig:test-info}). This is because the test unit will conduct multiple rounds of simulation tests. Considering the repetition of errors, we  only focus on the first error, which is Area 1, but there is some information here that does not have special guiding significance. As a result, the Input Area shown in the figure \ref{fig:test-info} is selected as the input of test information, including code fragments that fail to pass the test and error reporting results. This information is helpful for the model to determine the error type and error location, so as to guide the generation of higher quality code.

This paper analyzes the types of test information in the data set and their respective proportions. The test system will return a total of 7 types of test information, including pass (OK) and 6 types of errors. The error types are AssertionError, AttributeError, SyntaxError, NameError, TypeError  and IndentationError.

1. The AssertionError is shown in the test information in Figure \ref{fig:test-info}. The error is that the test system checks whether the code function is consistent with the preset function.

2. AttributeError, response object has no such method or class variable.

3. SyntaxError, which means that the code has a grammatical error, such as repeatedly defining parameters in a function definition.

4. NameError, indicating that the variable is not defined and used.

5. TypeErrors, indicating that the variable type is used inappropriately.

6. IndentatioErrors, indicating that the code format is wrong. 

The specific situation is shown in the appendix, and Table \ref{tab:Error-type} lists the proportion of each part of the output type of the test system in the training set.
\begin{table}[htbp]
  \centering
  \caption{Test information type and ratio}
    \begin{tabular}{p{9em}c}
    \toprule
    \textbf{ErrorType} & \multicolumn{1}{p{9em}}{\textbf{Proportion}} \\
    \midrule
    \textbf{OK} & 64.7 \\
    \textbf{AssertionError} & 14.2 \\
    \textbf{AttributeError} & 2.8 \\
    \textbf{SyntaxError} & 7.2 \\
    \textbf{NameError} & 9.3 \\
    \textbf{TypeError} & 0.9 \\
    \textbf{IndentationError} & 0.9 \\
    \bottomrule
    \end{tabular}%
  \label{tab:Error-type}%
\end{table}%

\subsection{Evaluation Metric and Model Setting}

\textbf{Test-Acc }This paper introduces a new metric, Test-Acc, which represents the rate of the generated code passing the test unit. Written the total number of samples as $M$, where the number of samples passing the test unit  is $N$ , then the Test-Acc is $\frac{N}{M}$. Test-Acc can compensate for shortcomings of Bleu\cite{papineni_bleu_2002} value, Rouge-L\cite{lin2006information}, StrAcc and ACC+\cite{sun_grammar-based_2019} in evaluating cases 2 and 3 described in \ref{sec:metric}. On the other hand, it can also verify whether the model can guide code generation after adding test information.

\textbf{Model Setting }The experiment was carried out on NVIDIA Tesla P4  and Ubuntu 18.04 LST system. The number of stacked blocks of Test-Info Encoder and Code Encoder were 6 and 5 respectively, and the number of iterations N was set to 3. Other parameters adopt the same settings as TreeGen \cite{sun_treegen_2019}. For example, the stacking number of NL Reader and AST Reader is 6 and 5 respectively, and the dimension size of word embedding is 256. For each fully connected layer, the dimension size is 256 except that the first layer is 1024 dimensions; dropout\cite{srivastava2014dropout} is applied after each layer (including attention layer, gating mechanism layer, convolutional layer and fully connected layer) with dropout rate of 0.15; Optimize through the default parameters of the Adafactor \cite{shazeer_adafactor_2018} optimizer.

\textbf{Training and Inference }The number of model iterations n is set to 1 to 3. The model is trained on the complete training set in the first round, in which the inputs of Test-Info Encoder and Code Encoder are empty. After the first round of model training, and we apply test process to output code. If some samples pass the test, there is no test error information.  We only select the samples with error information as the training samples of the next round, which ensures that the Test-Info Encoder and Code Encoder have information input, and let the model focus on learning the characteristics of error codes. Similarly, in the test set, we select the code with test information as the input to the model and get new output code , while the code without test information will be directly copied as the output of this round.

In the test set, according to the discriminate of the test unit for the output code in last round, directly copied  the code that passed the testing  as output code in the current round, and get Code Rule from the code that reported errors, NL , and test information  (Test-Info) as input for model to get new output code in the current round.

\section{result}
In order to evaluate the effectiveness of the automatic code generation model integrating test information, we study the following three questions.
\begin{itemize}
    \item RQ1: How does the performance of CodeGen-Test compare to other code generation models?
    \item RQ2: What is the impact of Test-Info and last code rule on the model?
    \item RQ3: How does the model perform under different iterations?
\end{itemize}

\subsection{How does the performance of CodeGen-Test compare to other code generation models?}

The model is compared with eight existing advanced models. The experimental results are shown in Table \ref{tab:result}. The comparison models are: (1) LPN, a potential predictor network \cite{ling_latent_2016}; (2) Seq2tree, a Neural attention model from language to logical form \cite{dong_language_2016}; (3) yin17, a Syntax neural model for general code generation \cite{yin_syntactic_2017}; (4) ASN, an Abstract syntax network model for code generation and semantic parsing \cite{rabinovich_abstract_2017}; (5) ReCode, Neural code generation model based on retrieval \cite{hayati_retrieval-based_2018}; (6) SZM, a grammar-based structural use CNN
decoder for code generation \cite{sun_grammar-based_2019}; (7) TreeGen, a transformer structure model for code generation \cite{sun_treegen_2019}; (8) ADG seq2seq, Neural code generation model embedded in API dependency graph \cite{lyu_embedding_2021}. Among the data of the comparison model, the Test-Acc results are calculated after we reproduce the model based on the open source code published by their respective authors. Rouge-L results are quoted from adg-seq2seq \cite{lyu_embedding_2021}, and other metric are quoted from the experimental results in their respective papers.
\begin{table}[htbp]
  \centering
  \caption{Model comparison experiments}
    \begin{tabular}{p{7em}lllll}
    \toprule
    Target & \multicolumn{1}{p{2.5em}}{\textbf{StrAcc}} & \multicolumn{1}{p{2.5em}}{\textbf{Acc+}} & \multicolumn{1}{p{2.5em}}{\textbf{Bleu}} & \multicolumn{1}{p{4em}}{\textbf{Rouge-L}} & \multicolumn{1}{p{4em}}{\textbf{Test-Acc}} \\
    \midrule
    LPN   & 6.1   & \multicolumn{1}{p{2.5em}}{-} & 67.1  & \multicolumn{1}{p{2.5em}}{-} & \multicolumn{1}{p{4em}}{-} \\
    SEQ2TREE & 1.5   & \multicolumn{1}{p{2.5em}}{-} & 53.4  & \multicolumn{1}{p{2.5em}}{-} & \multicolumn{1}{p{4em}}{-} \\
    YN17  & 16.2  & \multicolumn{1}{p{2.5em}}{18.2} & 75.8  & \multicolumn{1}{p{2.5em}}{-} & \multicolumn{1}{p{4em}}{-} \\
    ASN   & 18.2  & \multicolumn{1}{p{2.5em}}{-} & 77.6  & 77.5  & \multicolumn{1}{p{4.5em}}{-} \\
    ASN+supatt & 22.7  & \multicolumn{1}{p{2.5em}}{-} & 79.2  & \multicolumn{1}{p{2.5em}}{-} & \multicolumn{1}{p{4em}}{-} \\
    ReCode & 19.6  & \multicolumn{1}{p{2.5em}}{-} & 78.4  & \multicolumn{1}{p{2.5em}}{-} & \multicolumn{1}{p{4em}}{-} \\
    SZM19 & 27.3  & 30.3  & 79.6  & 82.8  & 33.3 \\
    TreeGen-A & 25.8  & 25.8  & 79.3  & \multicolumn{1}{p{2.5em}}{-} & \multicolumn{1}{p{4em}}{-} \\
    TreeGen-B & \textbf{31.8} & \textbf{33.3} & 80.8  & 82.9  & 37.9 \\
    ADG-Seq2Seq & 27.3  & \multicolumn{1}{p{2.5em}}{-} & 78.1  & 87.4  & \multicolumn{1}{p{4em}}{-} \\
    CodeGen-test & 28.8  & 28.8  & \textbf{81.0} & \textbf{87.7} & \textbf{43.9} \\
    \bottomrule
    \end{tabular}%
  \label{tab:result}%
\end{table}%

The results show that CodeGen-Test model has the best performance on Bleu, Rouge-L and Test-Acc metric. CodeGen-Test model outperform all methods except TreeGen-B in StrAcc and all methods except TreeGen-B and szm19 in ACC+ metric. This paper argues that the Test-Acc is more important than StrAcc and Acc+, because StrAcc and Acc+ 
reflect the extent that the generated code same  with the reference code at the character level . However,  different from natural language, we  not only pay attention to the similarity of code at the character level, but also pay more attention to the  correctness of code  function. The test accuracy rate (Test-Acc) metric can reflect the correctness of the code's grammar and function. The above experimental results verify the effectiveness of CodeGen-Test in the task of automatic code generation.

\subsection{What is the impact of Test-Info and last code rule on the model?}
In order to verify this problem, the Test-Info Encoder and Code Encoder in the model are first removed respectively, and then completely, so as to explore the improvement performance of the test information and the code generated by the previous round of model on the model. In addition, comparing the model with different iteration times can also verify the problem from another perspective, when the iteration times n > 1, that is, when the test information and the code generated in the previous round are added, compared with the iteration times n = 1, that is, when the test information and the code generated in the previous round are not added, whether the model performance is improved.

The experimental results are shown in Table \ref{tab:result2}. It can be seen that the model achieves the best results under the complete model, and the sequential removal and total removal of Test-Info Encoder and Code Encoder have a negative impact on the model performance. In addition, in the case of a complete model, adding test information and the code generated in the previous round (number of iterations N>1), comparing the code generated in the previous round without adding test information (number of iterations N=1), the model performance is also promoted.

To sum up the experimental results, it can be concluded that both Test-Info and last code rule can improve the model, and the best performance can be achieved when they are input into the model at the same time.

\subsection{How does the model perform under different iterations?}
In order to verify the results of this problem, experiments are carried out with the number of iterations n from 1 to 3.
\begin{table}[htbp]
  \centering
  \caption{Ablation experiments}
    \begin{tabular}{cccccc}
    \toprule
    \multicolumn{1}{p{2em}}{\textbf{Model}} & \multicolumn{1}{p{2.375em}}{\textbf{N}} & \multicolumn{1}{p{2.375em}}{\textbf{StrAcc}} & \multicolumn{1}{p{2.375em}}{\textbf{Acc+}} & \multicolumn{1}{p{2.375em}}{\textbf{Bleu}} & \multicolumn{1}{p{2.375em}}{\textbf{Test-Acc}} \\
    \midrule
    \multicolumn{1}{l}{\multirow{3}[2]{*}{\shortstack{CodeGen-Test\\ (full model)}}} & 1     & 24.2  & 24.2  & 78.9  & 37.9 \\
          & 2     & 28.8  & 28.8  & 81.0    & 43.9 \\
          & 3     & 28.8  & 28.8  & 81.0    & 43.9 \\
    \midrule
    \multicolumn{1}{l}{\multirow{3}[2]{*}{\shortstack{CodeGen-Test \\ (w/o Test-Info Encoder)}}} & 1     & 27.3  & 27.3  & 78.9  & 37.9 \\
          & 2     & 27.3  & 27.3  & 79.7  & 37.9 \\
          & 3     & 27.3  & 27.3  & 79.7  & 40.9 \\
    \midrule
    \multicolumn{1}{l}{\multirow{3}[2]{*}{\shortstack{CodeGen-Test \\ (w/o Code Encoder)}}} & 1     & 22.7  & 22.7  & 77.5  & 31.8 \\
          & 2     & 25.8  & 25.8  & 79.2  & 34.8 \\
          & 3     & 27.3  & 27.3  & 79.8  & 37.9 \\
    \midrule
    \multicolumn{1}{l}{\multirow{3}[2]{*}{\shortstack{CodeGen-test \\ (w/o Test-Info Encoder ,\\  Code Encoder)}}} &  1 & 22.7  & 22.7  & 78.5  & 34.8 \\
          & 2     & 24.2  & 24.2  & 79.0    & 37.9 \\
          & 3     & 24.2  & 24.2  & 79.2  & 37.9 \\
    \bottomrule
    \end{tabular}%
  \label{tab:result2}%
\end{table}%

The experimental results are shown in Table \ref{tab:result2}. On the one hand, the model performance is improved with the increase of iteration times. The results of the four models in Table \ref{tab:result2} are better when iteration times n = 3 than  iteration times n = 1; On the other hand, it is also noted that when n $\ge$ 2, the increase of iteration times does not significantly  improve the model. In Table \ref{tab:result2}, only the third and fourth models are improved with the increase of iteration times when iteration times n $\ge$ 2, while the performance of the first and second models have stagnated when iteration times n $\ge$ 2. Besides, during the experiment, 
we found that when n = 3, most models can not get a better model in the verification set during training, so they are not improved in the test set.

We think this is partly due to the gradual decrease in the number of data sets during the experiment. We lists the number of samples in each iteration of CodeGen-Test model in Table \ref{tab:train set number}. The total number of samples in the training set is only 533. With the increase of the number of iterations n, the number of samples with test information content is further reduced, which limits the further improvement of the model.

Overall, the model performance improves with  iteration times increase, but when n $\ge$ 2, the model performance does not improve significantly with iteration times increase, We think this is partly due to the gradual decrease in the number of data sets during the experiment.
\begin{table}[htbp]
  \centering
  \caption{Number of samples in data sets of each round}
    \begin{tabular}{lll}
    \toprule
    \multicolumn{1}{p{5em}}{\textbf{N}} & \multicolumn{1}{p{5em}}{\textbf{Train}} & \multicolumn{1}{p{5em}}{\textbf{Dev}} \\
    \midrule
    1     & 533   & 66 \\
    2     & 166   & 39 \\
    3     & 104   & 35 \\
    \bottomrule
    \end{tabular}%
  \label{tab:train set number}%
\end{table}%

\section{conclusion}
This paper proposes an automatic code generation model, CodeGen-Test that integrates program test information. The model is based on sequence to sequence framework. In this paper, we also test the generated code, improve  the process of automatic code generation technology, and integrate  program test information to guide the model. The model was tested on the "Heartstone" data set, and the results showed that the CodeGen-Test model has a better performance on each evaluation metric than other models, reflecting the effectiveness of the model. This paper also proposes a new evaluation metric, the Test-Acc of the code, compared with the existing evaluation metric, it can measure the quality of code  in terms of code function .

The model has a certain improvement in the preformance of automatic code generation tasks, mainly because of the integration of test information. Therefore, the model needs to obtain test information and multiple rounds of training, which increases the experimental steps and time. In the future work, We hope to improve the model from the following two aspects: First, we intend to combine the information of the code API to guide the model to further improve in the case of function call and parameter transfer. Second, the information in similar requirements and codes could be directly used when generating codes by searching similar language code pairs. Thus, the possibility of generating wrong codes could be reduced.

\ifCLASSOPTIONcompsoc
  \section*{Acknowledgments}
\else
  \section*{Acknowledgment}
\fi

The research is supported by the National Natural Science Foundation (NNSF) of China (No. 61877031, No. 61876074, No. 61966019, No. 61977038). We thank all reviewers for their valuable comments.

\ifCLASSOPTIONcaptionsoff
  \newpage
\fi



\bibliographystyle{IEEEtran}
\bibliography{ref.bib}
%




%

\begin{IEEEbiography}[{\includegraphics[width=1in,height=1.25in,clip,keepaspectratio]{ 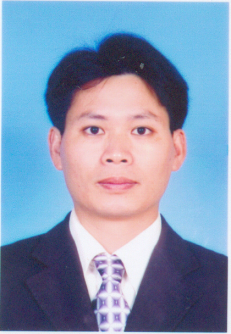}}]{ZHONG Maosheng}
  is currently a Full Professor with Jiangxi Normal University, Nanchang, Jiangxi, China. He has published more than 80 international journal or conference papers. His research interests include machine learning, software engineering and natural language processing. He received his Ph.D. degree in computer software and theory from Shanghai JiaoTong University in 2010. 
\end{IEEEbiography}

\begin{IEEEbiography}[{\includegraphics[width=1in,height=1.25in,clip,keepaspectratio]{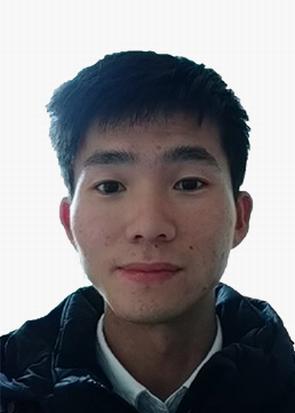}}]{LIU Gen}
LIU Gen is currently a graduate student at Jiangxi Normal University, NanChang, Jiangxi, China. He  research interests include code summarization and code generation. 
\end{IEEEbiography}


\begin{IEEEbiography}[{\includegraphics[width=1in,height=1.25in,clip,keepaspectratio]{ 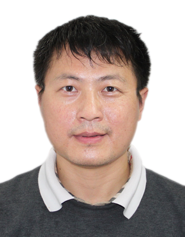}}]{LI Hongwei}
  born in 1975. PhD. Associate Professor. Member of China Computer Federation. He received his PhD degree of computer science and technology from Fudan University, Shanghai in 2016. His main research interests include program analysis and empirical software engineering, software knowledge graph, natural language processing.
\end{IEEEbiography}

\begin{IEEEbiography}[{\includegraphics[width=1in,height=1.25in,clip,keepaspectratio]{ 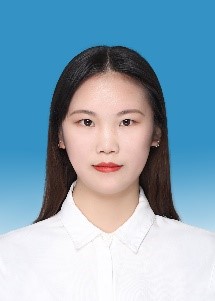}}]{KUANG Jiangling}
  is currently a graduate student at Jiangxi Normal University, NanChang, Jiangxi, China. Her research interests include code summarization and code generation. She received her B.S degree at the School of Software Engineering.
\end{IEEEbiography}

\begin{IEEEbiography}[{\includegraphics[width=1in,height=1.25in,clip,keepaspectratio]{ 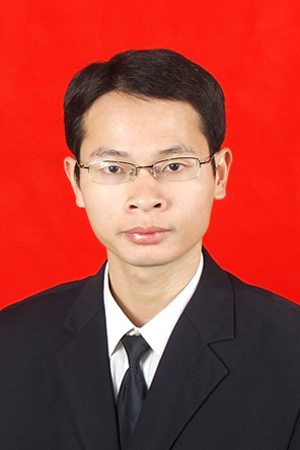}}]{ZENG Jinshan}
received the PhD degree in mathematics from Xi'an Jiaotong University, Xi'an, China, in 2015. He is currently a distinguished professor with the School of Computer and Information Engineering, Jiangxi Normal University, Nanchang, China, and serves as the director of the department of data science and big data. He has published over 40 papers in high-impact journals and conferences such as IEEE TPAMI, JMLR, IEEE TSP, ICML, AAAI etc. He has had two papers co-authored with collaborators that received the International Consortium of Chinese Mathematicians (ICCM) Best Paper Award (2018, 2020). His research interests include nonconvex optimization, machine learning (in particular deep learning), and remote sensing.
\end{IEEEbiography}

\begin{IEEEbiography}[{\includegraphics[width=1in,height=1.25in,clip,keepaspectratio]{ 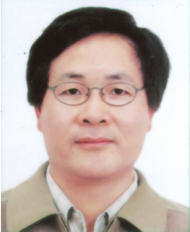}}]{WANG Mingwen}
  received his PhD degree in computer software and theory from
Shanghai Jiao Tong University in 2000. He is currently a professor and a supervisor of doctor candidates in School of Computer and Information Engineering, Jiangxi Normal University, Nanchang, China. He has published more than 100 international journal or conference papers. His research interests include information retrieval, data mining, software engineering and natural language processing.
\end{IEEEbiography}


\vfill

\begin{appendices}
\clearpage
\appendices
\section{Examples of types of test error information}


\begin{figure}[htbp]
\centering
\subfigure[Error Code]{
    \begin{minipage}[b]{0.5\textwidth}
    \includegraphics[width=1\textwidth]{ 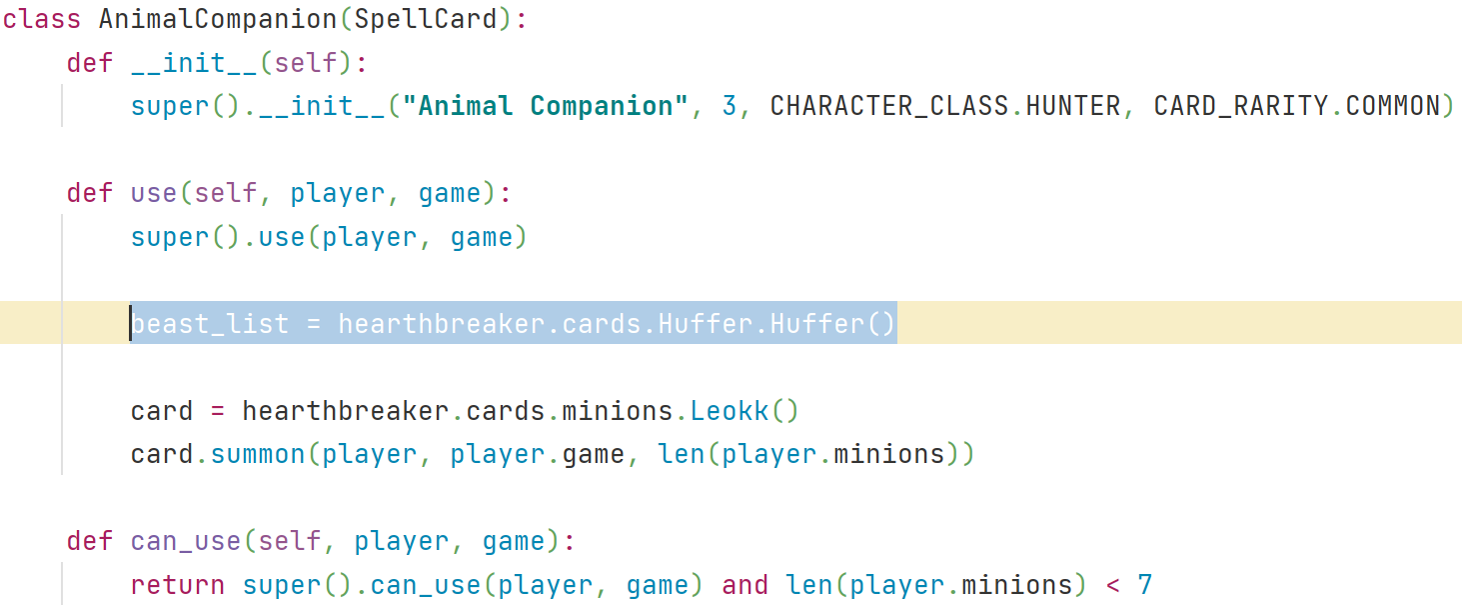}
    \end{minipage}
}

\subfigure[Test information]{
    \begin{minipage}[b]{0.5\textwidth}
    \includegraphics[width=1\textwidth]{ 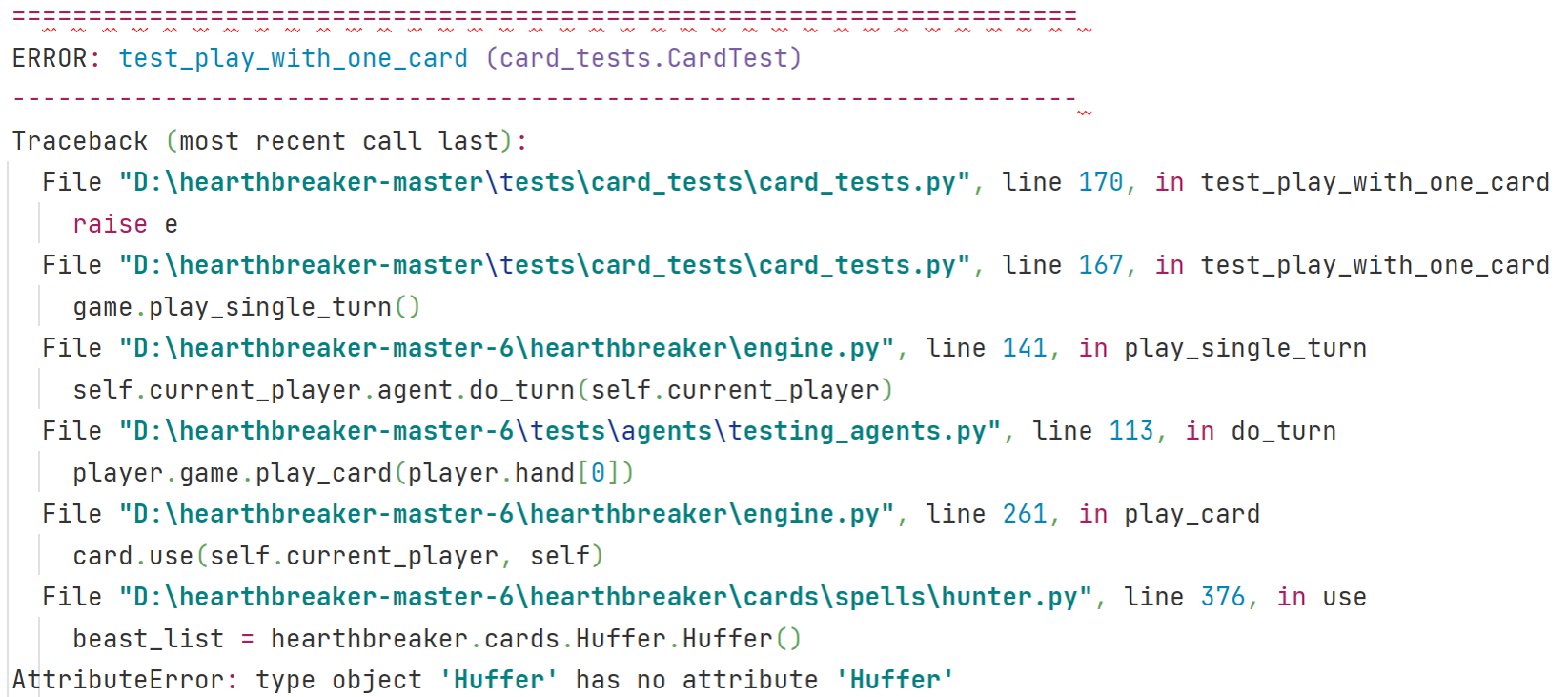}
    \end{minipage}
}
\caption{Examples of AttributeError codes and test information}
\end{figure}


\begin{figure}[htbp]
\centering
\subfigure[Error Code]{
    \begin{minipage}[b]{0.5\textwidth}
    \includegraphics[width=1\textwidth]{ 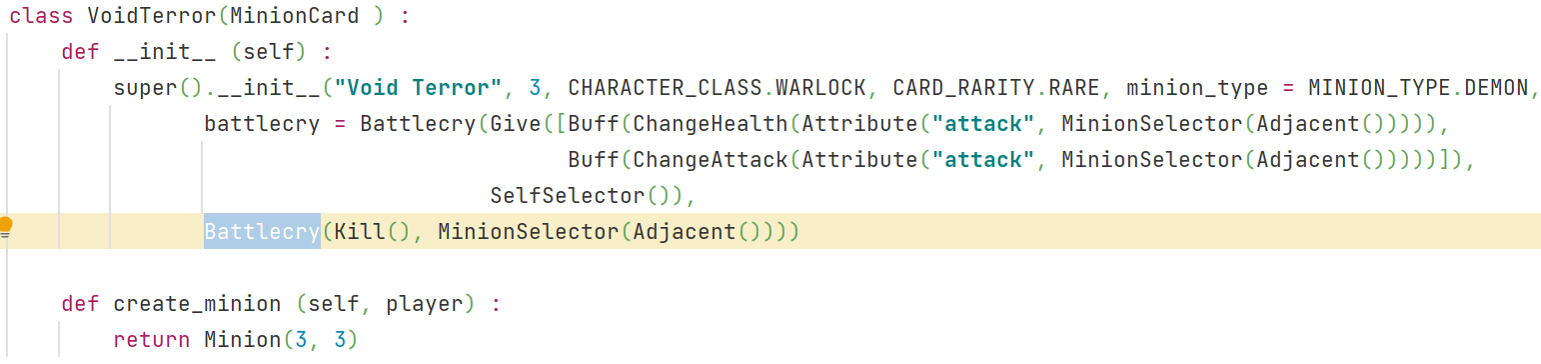}
    \end{minipage}
}

\subfigure[Test information]{
    \begin{minipage}[b]{0.5\textwidth}
    \includegraphics[width=1\textwidth]{ 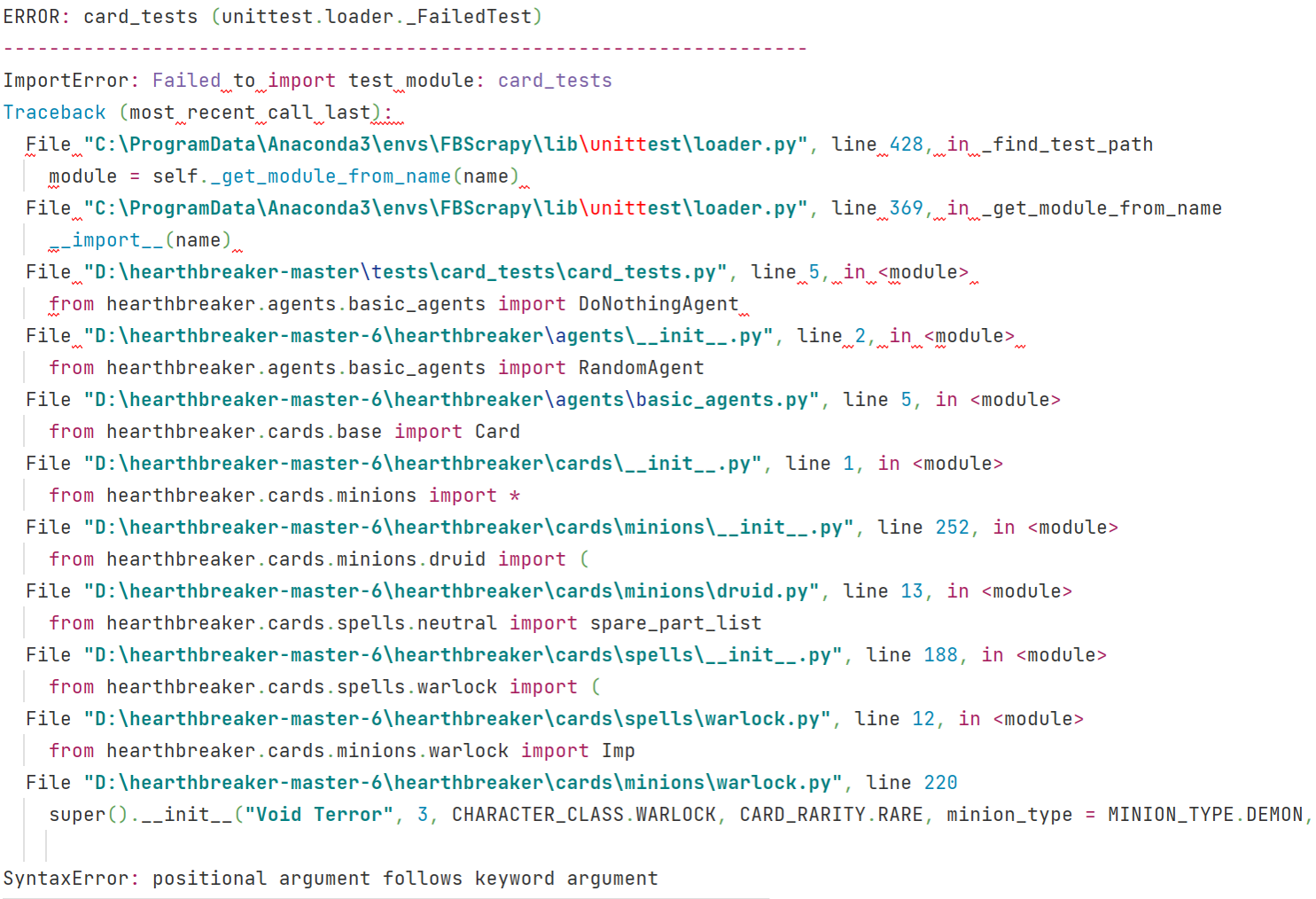}
    \end{minipage}
}
\caption{Examples of SyntaxError codes and test information}
\end{figure}


\begin{figure}[htbp]
\centering
\subfigure[Error Code]{
    \begin{minipage}[b]{0.5\textwidth}
    \includegraphics[width=1\textwidth]{ 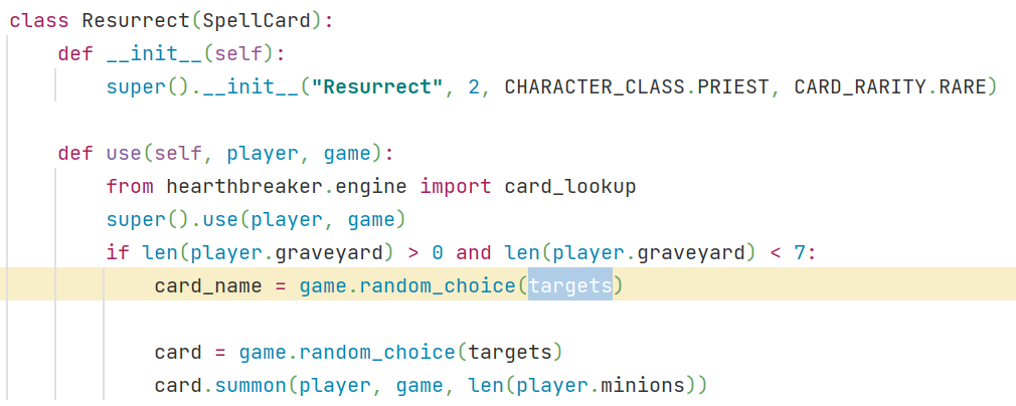}
    \end{minipage}
}

\subfigure[Test information]{
    \begin{minipage}[b]{0.5\textwidth}
    \includegraphics[width=1\textwidth]{ 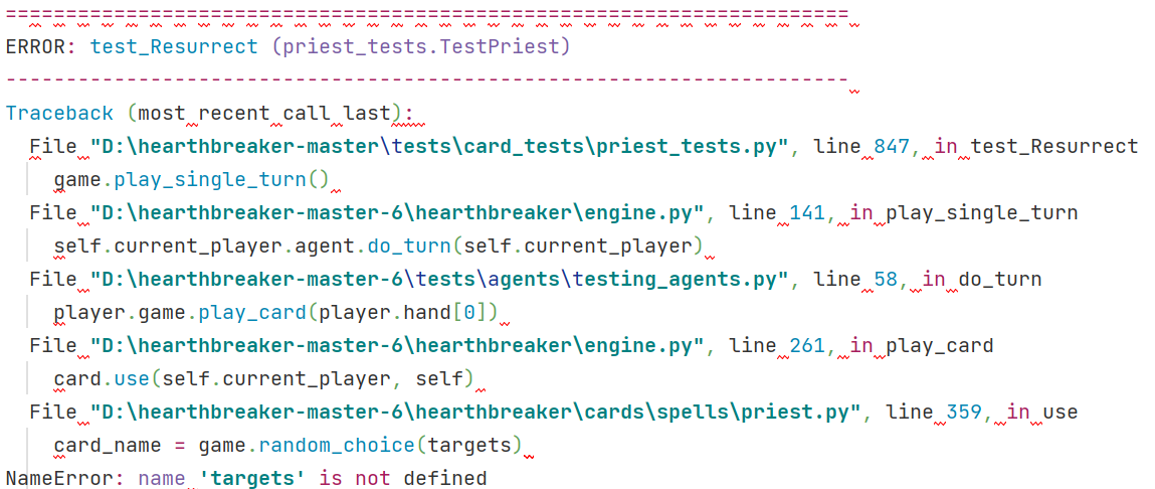}
    \end{minipage}
}
\caption{Examples of NameError codes and test information}
\end{figure}


\begin{figure}[htbp]
\centering
\subfigure[Error Code]{
    \begin{minipage}[b]{0.5\textwidth}
    \includegraphics[width=1\textwidth]{ 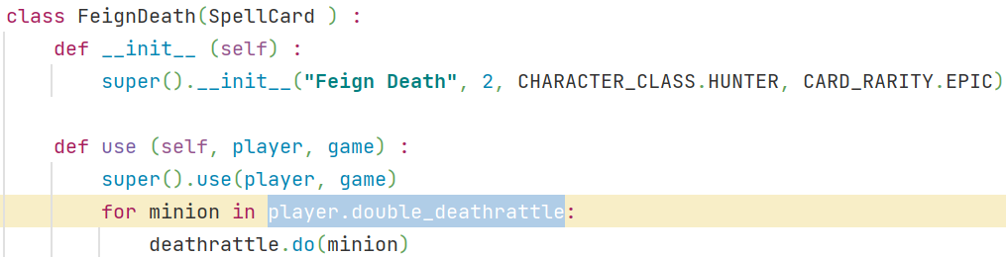}
    \end{minipage}
}

\subfigure[Test information]{
    \begin{minipage}[b]{0.5\textwidth}
    \includegraphics[width=1\textwidth]{ 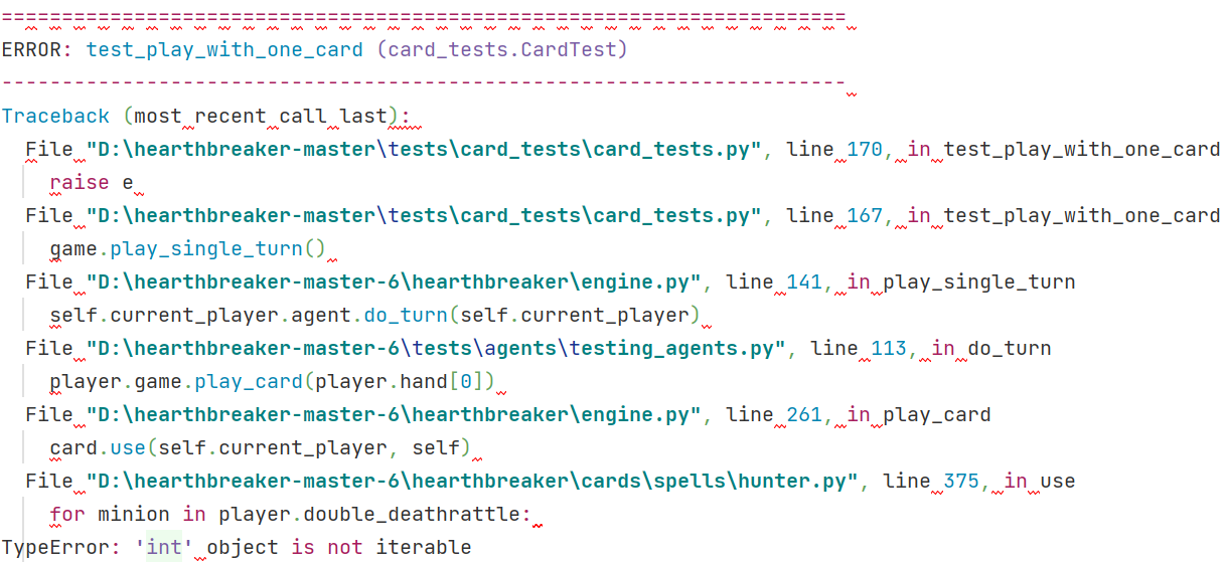}
    \end{minipage}
}
\caption{Examples of TypeError codes and test information}
\end{figure}


\begin{figure}[htbp]
\centering
\subfigure[Error Code]{
    \begin{minipage}[b]{0.5\textwidth}
    \includegraphics[width=1\textwidth]{ 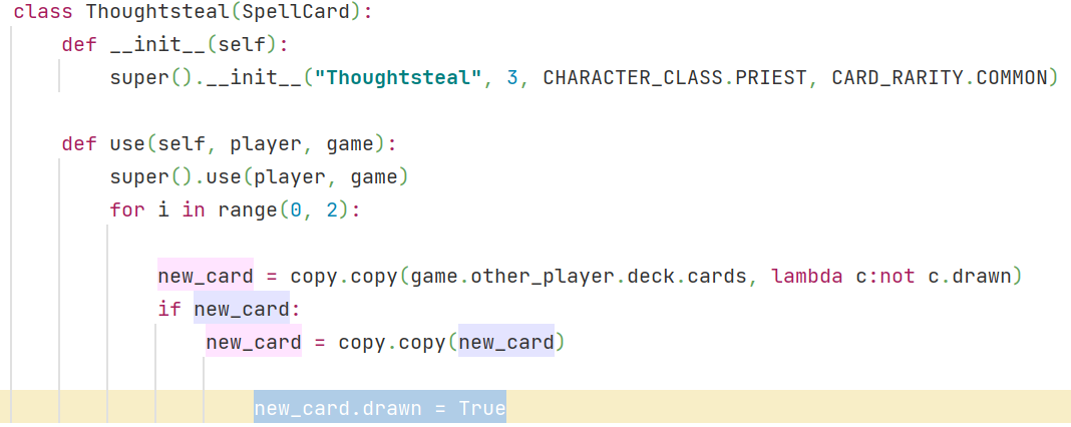}
    \end{minipage}
}

\subfigure[Test information]{
    \begin{minipage}[b]{0.5\textwidth}
    \includegraphics[width=1\textwidth]{ 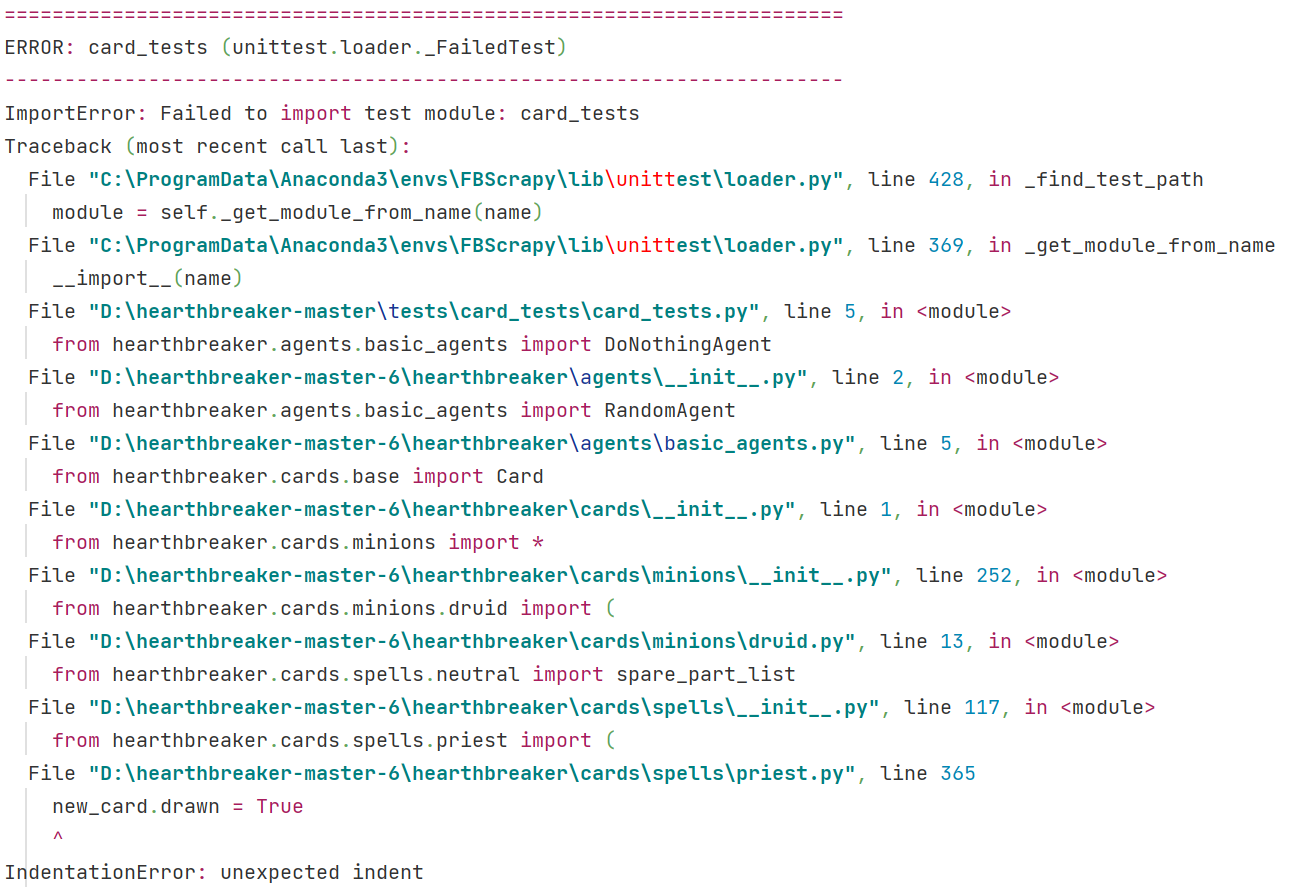}
    \end{minipage}
}
\caption{Examples of IndentationError codes and test information}
\end{figure}
\end{appendices}

\end{document}